\documentclass[draft]{agujournal2019}
\usepackage{url} %this package should fix any errors with URLs in refs.
\usepackage{lineno}
\usepackage{soul}
\usepackage{amsmath}
\usepackage{csquotes}
\usepackage{booktabs}
\usepackage{MnSymbol}
\usepackage{amsfonts}

\usepackage{pdflscape}

\DeclareMathOperator{\cmss}{cmss}

\drafttrue

\journalname{JGR: Oceans}

\begin{document}

%% ------------------------------------------------------------------------ %%
%  Title
%
% (A title should be specific, informative, and brief. Use
% abbreviations only if they are defined in the abstract. Titles that
% start with general keywords then specific terms are optimized in
% searches)
%
%% ------------------------------------------------------------------------ %%

\title{A New Probabilistic  Wave Breaking Model for Dominant Wind-sea Waves Based on the Gaussian Field Theory}

%% ------------------------------------------------------------------------ %%
%
%  AUTHORS AND AFFILIATIONS
%
%% ------------------------------------------------------------------------ %%

% Authors are individuals who have significantly contributed to the
% research and preparation of the article. Group authors are allowed, if
% each author in the group is separately identified in an appendix.)

% List authors by first name or initial followed by last name and
% separated by commas. Use \affil{} to number affiliations, and
% \thanks{} for author notes.
% Additional author notes should be indicated with \thanks{} (for
% example, for current addresses).

% Example: \authors{A. B. Author\affil{1}\thanks{Current address, Antartica}, B. C. Author\affil{2,3}, and D. E.
% Author\affil{3,4}\thanks{Also funded by Monsanto.}}

\authors{C. E. Stringari\affil{1}, M. Prevosto\affil{2}, J.-F. Filipot\affil{1}, F. Leckler\affil{1}, P. V. Guimar\~aes\affil{1,3}}

\affiliation{1}{France Energies Marines, Plouzané, France}
\affiliation{2}{Institut Français de Recherche pour l'Exploitation de la Mer, Plouzané, France}
\affiliation{3}{PPGOceano, Federal University of Santa Catarina, Florianópolis, 88040-900, Brazil}

% \affiliation{2}{Second Affiliation}
% \affiliation{3}{Third Affiliation}
% \affiliation{4}{Fourth Affiliation}

%\affiliation{=number=}{=Affiliation Address=}
%(repeat as many times as is necessary)

%% Corresponding Author:
% Corresponding author mailing address and e-mail address:

% (include name and email addresses of the corresponding author.  More
% than one corresponding author is allowed in this LaTeX file and for
% publication; but only one corresponding author is allowed in our
% editorial system.)

% Example: \correspondingauthor{First and Last Name}{email@address.edu}
\correspondingauthor{C.E. Stringari}{Caio.Stringari@france-energies-marines.org}
\correspondingauthor{J.F. Filipot}{Jean.Francois.Filipot@france-energies-marines.org}

%% Keypoints, final entry on title page.

%  List up to three key points (at least one is required)
%  Key Points summarize the main points and conclusions of the article
%  Each must be 100 characters or less with no special characters or punctuation and must be complete sentences

% Example:
% \begin{keypoints}
% \item	List up to three key points (at least one is required)
% \item	Key Points summarize the main points and conclusions of the article
% \item	Each must be 140 characters or less with no special characters or punctuation and must be complete sentences
% \end{keypoints}

\begin{keypoints}
\item A new probabilistic wave breaking model based on Gaussian field theory is presented for dominant, wind-sea waves.

\item Wave breaking probabilities are modeled from the joint probability density between wave phase speed and particle orbital velocity.

\item The proposed model performs well when compared to six other historical models using three field datasets.

\end{keypoints}

%% ------------------------------------------------------------------------ %%
%
%  ABSTRACT and PLAIN LANGUAGE SUMMARY
%
% A good Abstract will begin with a short description of the problem
% being addressed, briefly describe the new data or analyses, then
% briefly states the main conclusion(s) and how they are supported and
% uncertainties.

%
%% ------------------------------------------------------------------------ %%

%% \begin{abstract} starts the second page

\begin{abstract}

 This paper presents a novel method for obtaining the probability wave of breaking ($P_b$) of deep water, dominant wind-sea waves (that is, waves made of the energy within $\pm$30\% of the peak wave frequency) derived from Gaussian wave field theory. For a given input wave spectrum we demonstrate how it is possible to derive a joint probability density function between wave phase speed ($c$) and horizontal orbital velocity at wave crest ($u$) from which a model for $P_b$ can be obtained. A non-linear kinematic wave breaking criterion consistent with the Gaussian framework is further proposed. Our model would allow, therefore, for application of the classical wave breaking criterion (that is, wave breaking occurs if $u/c > 1$) in spectral wave models which, to the authors' knowledge, has not been done to date.  Our results show that the proposed theoretical model has errors in the same order of magnitude  as six other historical models when assessed using three field datasets. With optimization of the proposed model's single free parameter, it can become the best performing model for specific datasets. Although our results are promising, additional, more complete wave breaking datasets collected in the field are needed to comprehensively assess the present model, especially in regards to the dependence on phenomena such as direct wind forcing, long wave modulation and wave directionality. 

\end{abstract}

\section*{Plain Language Summary}

% Please make you sure you read this before editing this section:

% The Plain Language Summary should be written for a broad audience, including journalists and the science-interested public, that will not have a background in your field. A Plain Language Summary is required in GRL, JGR: Planets, JGR: Biogeosciences, JGR: Oceans, G-Cubed, Reviews of Geophysics, and JAMES. see http://sharingscience.agu.org/creating-plain-language-summary/)

% Absolutely no jargon allowed in this section. Please do not use "dynamics", "phase speed", "field", "celerity", "velocity x speed", etc. This section must be easy to read and is not target at specialists. I am very likely to reject your changes to this section so if unsure on what you are writing, please ask me before making any changes.

Waves will break if the speed of the water particles on the wave crest is greater than the speed of the wave itself, causing the wave crest to overtake the front part of the wave, leading to wave breaking. Precisely simulating real ocean waves requires, therefore, a particle-by-particle description of the water motion, which is too expensive for the current computers to handle in real-world applications. Instead, wave models describe waves by means of their statistical properties, that is, averaged over a large number of waves. In this paper, we present a mathematical formulation that allows to calculate the combined probability between the speed of particles on the wave crest and the wave speed based only on statistical properties. From these combined probabilities, we model the probability of wave breaking. Our results indicate that our model performed relatively well when compared to six other models using three historical datasets. Because of a lack of observed data to assess our model, we recommend that future research should focus on collecting more wave breaking data measured in the field. Future advances on this line of research could lead, for example, to improvements on operational weather forecast models.

%% ------------------------------------------------------- %%
%
%  TEXT
%
%% ------------------------------------------------------- %%

\section{Introduction}\label{sec:intro}

A robust description of wave breaking is a crucial aspect of wave modelling. It is via wave breaking that most of the wave energy is dissipated and a precise formulation of this phenomenon is required to obtain reliable models. Despite of its importance, energy dissipation due to wave breaking is still modelled as a semi-empirical process due to the difficulty to represent physically-derived wave breaking criteria on phase-averaged wave models \cite{Battjes1978, Thornton1983, Banner2000, Filipot2010, Filipot2012, Banner2002, Ardhuin2010, Banner2014, Zieger2015,Ardag2020}. The available probabilistic (that is, parametric, or empirical) formulations included in these models have been derived from limited datasets and without rigorous theoretical frameworks and, therefore, they currently lack a solid physical background. While the current operational (spectral) models are capable of reproducing field observations of integrated spectral parameters (for example, significant wave height, peak wave period and peak wave direction) with good accuracy, it remains unclear if their wave breaking parameterizations are entirely reliable. This knowledge gap partly occurs because limited research has focused on wave breaking statistics derived from field data, especially when it comes to wave breaking observations distributed as a function of wave scales (for example, wave frequency or wave phase speed). The research developed here has, therefore, important implications for air-sea flux parameterizations \cite{Kudryavtsev2014}, safety at sea \cite{Kjeldsen1980} and design of offshore structures \cite{Filipot2019}, all of which directly rely on the properties of breaking waves.

Historically, parametric wave breaking models have been constructed from two different approaches: the first approach considers wave statistics (wave steepness, most frequently) derived from a wave-by-wave analysis of the surface elevation timeseries collected at a single point location where wave breaking occurrences are synchronously identified (using video data, most frequently). The wave breaking probability (that is, the ratio between the total number of breaking waves over the total number of waves during a given period of time) can then be expressed as a bulk quantity \cite{Thornton1983, Chawla&Kirby2002, Alsina2007, Janssen2007} or can be distributed over wave frequency ($f$), wavenumber ($k$), or wave speed ($c$) ranges, referred as to ``wave scales'' by the wave modelling community \cite{Eldeberky1996, Banner2002, Filipot2010}. 

The second approach follows from \citeA{Phillips1985} who defined the distribution $\Lambda(c)dc$ as the \textquote{average total length per unit surface area of breaking fronts that have velocities in the range $c$ to $c+dc$}. This approach therefore relates to the analysis of sea surface images in which individual wave breaking patches are tracked in space and time. The main motivation for introducing this new concept was clearly stated in \citeA{Phillips1985}: \enquote{There is clearly some association of the breaking events with waves of different scales, but it is difficult to make the association in an unambiguous way if we consider only the surface configuration at one given instant. A breaking crest may indeed be a local maximum in the instantaneous surface configuration but there is no guarantee that a local wavelength of the breaking wave can be defined clearly. It seems more satisfactory to use the velocity $c$ of the breaking front as a measure of the scale of the breaking}. This quotation clearly identify the limitations of directly relying on the analysis of single point elevation timeseries. Different parameterizations have been proposed to quantity $\Lambda(c)dc$ from theoretical \cite{Phillips1985} or empirical considerations \cite{Melville2002, Sutherland2013, Romero2019}. However, \citeauthor{Phillips1985}' \citeyear{Phillips1985} framework remains controversial, particularly regarding its practical application, given that different interpretations of his concepts can generate differences of several orders of magnitude in the calculations of $\Lambda(c)dc$ and its moments \cite{Banner2014}. For a detailed review of commonly used parametric wave breaking models please refer to \ref{sec:review:parametric:models}.

Interestingly, while the ratio between the horizontal orbital velocity at the crest ($u$) to wave phase speed ($c$) appears the most reliable parameter to determine wave breaking occurrence \cite{Saket2017, Barthelemy2018, Derakhti2020, Varing2020}, it was not used by any of the approaches mentioned above. This paper provides a new promising wave breaking model by revisiting \citeA{Rice1944} and \citeA{Longuet-Higgins1957} statistical descriptions of Gaussian processes (that is, for linear waves) to obtain the theoretical joint probability density between $c$ and $u$ ($p(c,u)$). We then model $P_b$ assuming a kinematic wave breaking criterion consistent with non-linear waves, that is, a wave breaks if the fluid velocity at the wave crest is greater than the wave phase speed ($u$ $>$ $c$). This study focuses on analysing dominant waves, defined as waves that have frequencies within $\pm$30\% of the spectral peak frequency of the wind-sea \cite{Banner2000}. Future research will be dedicated to extend our efforts to broader wave scales. This paper is organized as follows: Section \ref{sec:gaussian:waves} describes the proposed model, Section \ref{sec:materials:methods:field:data} presents three historical datasets used to evaluate the model, Section \ref{sec:results} presents the results, Section \ref{sec:discussion} discusses and Section \ref{sec:conclusion} concludes.

\section{Definition of a Probabilistic Wave Breaking Model Based on Gaussian Field Theory}\label{sec:gaussian:waves}

The kinematic wave breaking criterion $u/c=1$ has been historically used as the onset of wave breaking for non-linear, real waves (see \citeA{Perlin2013} for a review). Recently, \citeA{Barthelemy2018} found and \citeA{Derakhti2020} confirmed via numerical simulations that waves will inevitably start to break shortly after $u/c$ exceeds 0.85 in deep and shallow water. Further numerical simulations showed that wave breaking occurs when the maximum orbital velocity ($u_{max}$) equals $c$ somewhere along the wave  profile and not necessarily at the wave crest \cite{Varing2020}. Although the relationship $u/c$ provides a solid physical background to establish the onset of wave breaking, this approach has never been applied to spectral wave models because it requires phase-resolving the wave field. In the sections below, we circumvent this difficulty by defining a wave breaking probability model using the joint probability density between $c$ and $u$ corresponding to a given wave energy spectrum ($E(f)$). The efforts in this paper are consistent with part of the recent work from \citeA{Ardag2020} in the sense that both works aim to solidify the use of the kinematic wave breaking criterion as the standard approach for modelling wave breaking.

\subsection{Theoretical Derivation of the Joint Probability Density Distribution of Orbital Velocity at the Wave Crest and Phase Speed}\label{sec:gaussian:waves:1}

\citeA{Longuet-Higgins1957} published a very complete work on the statistics of Gaussian wave fields. In particular, \citeA{Longuet-Higgins1957} studied the probability density of the speed of zero-crossings along a given line that is of interest for us in this work. In his paper, the speed of zero-crossings were applied in particular to the zero-crossings of the space derivative of a Gaussian process, that is, the velocities of the local maxima in space (\citeA{Longuet-Higgins1957}, pp. 356-357). The present work describes how the same methodology can be extended to derive the joint density of the speed of space local maxima (or local crests) and simultaneous wave horizontal orbital velocity for a one-dimensional Gaussian sea state. For simplicity, this paper follows the same notations as those of \citeA{Longuet-Higgins1957} and the reader is directed to Section 2.5 in \citeA{Longuet-Higgins1957} for further details.

As explained in \citeA{Longuet-Higgins1957}, if $\xi_{1}\left(x,t\right)$  is a stationary-homogeneous process and we are interested in the points (for example, in space) were this process crosses a level $x_{1}$, the joint distribution of the space derivative of $\xi_{1}$ noted $\xi_2$, with other related processes $\xi_{3}$, $\xi_{4}$, $\ldots$ at $\xi_1 = x_1$ is given by:

\begin{equation}
p\left(\xi_{2},\xi_{3},\xi_{4},...\right)_{x_{1}}=\frac{\left.\left|\xi_{2}\right|p\left(\xi_{1},\xi_{2},\xi_{3},\xi_{4},...\right)\right|_{\xi_{1}=x_{1}}}{N_{0}\left(x_{1}\right)}\label{eq:Rice}
\end{equation}

\noindent where $N_{0}\left(x_{1}\right)$ is the number of crossings of the
level $x_{1}$ by $\xi_{1}$ (see Equation 2.2.5 in \citeA{Longuet-Higgins1957}). In this paper we are interested in joint distributions at the local maxima in space of the wave elevation process $\xi_{0}$. Therefore, $\xi_{1}$ is the space derivative of the wave process and local maxima correspond to down-crossings of the zero level by $\xi_{1} =\partial\xi_{0}/\partial x$.

\begin{equation}
\xi_{1}=\frac{\partial\xi_{0}}{\partial x}\;,\;\xi_{2}=\frac{\partial^{2}\xi_{0}}{\partial x^{2}}=\frac{\partial\xi_{1}}{\partial x}.
\end{equation}

\noindent In the case of Gaussian processes, $N_{0}^{-}\left(x_{1}\right)$ is:

\begin{equation}
N_{0}^{-}\left(x_{1}\right)=\frac{1}{2\pi}\sqrt{\frac{m_{4}}{m_{2}}}\exp\left(-\frac{x_{1}^{2}}{2m_{2}}\right)\;,\;N_{0}^{-}=N_{0}^{-}\left(0\right)=\frac{1}{2\pi}\sqrt{\frac{m_{4}}{m_{2}}}\label{eq:N_0}
\end{equation}

\noindent  where $m_0, \ m_1, \ \ldots, \ m_{i}$ are the $i$-th wavenumber spectral moments and the minus sign indicates that we consider only down-crossings.

\subsubsection{Speed of Local Maxima (Phase Speed)}

Following \citeA{Longuet-Higgins1957}, if we are interested in the speed $c$ of the local maxima in space, that is, the speed of the down-crossings of $\xi_{1}$, we have:

\begin{equation}
c=-\frac{\left.\partial\xi_{1}\right/\partial t}{\left.\partial\xi_{1}\right/\partial x}=-\frac{\xi_{3}}{\xi_{2}}\;\mathrm{with}\;  \xi_{2}=\frac{\partial^{2}\xi_{0}}{\partial x^{2}} \mathrm{ and }\;\xi_{3}=\left.\partial\xi_{1}\right/\partial t .
\end{equation}

\noindent Using Equation \ref{eq:Rice},

\begin{equation}
p\left(\xi_{2},\xi_{3}\right)_{0}=\frac{\left.\left|\xi_{2}\right|p\left(\xi_{1},\xi_{2},\xi_{3}\right)\right|_{\xi_{1}=0}}{N_{0}^{-}}
\end{equation}

\noindent with $p\left(\xi_{1},\xi_{2},\xi_{3}\right)$ the point joint distribution of the three Gaussian processes $\frac{\partial\xi_{0}}{\partial x},\frac{\partial^{2}\xi_{0}}{\partial x^{2}},\frac{\partial^{2}\xi_{0}}{\partial x\partial t}$ is:

\begin{equation}
p\left(\xi_{1},\xi_{2},\xi_{3}\right)=p\left(\xi_{1}\right)p\left(\xi_{2},\xi_{3}\right)=\frac{{\rm e}^{-\frac{\xi_{1}^{2}}{2m_{2}}}}{2\pi\sqrt{m_{2}}}\frac{{\rm e}^{-\frac{1}{2}\left(\left[\xi_{2}\xi_{3}\right]Q_{c}^{-1}\left[\begin{array}{c}
\xi_{2}\\
\xi_{3}
\end{array}\right]\right)}}{\sqrt{\left(2\pi\right)^{3}\det(Q_{c})}}
\end{equation}

\noindent and covariance matrix:

\begin{equation}
\label{eq:Q}
Q=\left[\begin{array}{ccc}
m_{2} & 0 & 0\\
0 & m_{4} & m_{3}^{'}\\
0 & m_{3}^{'} & m_{2}^{''}
\end{array}\right]=\left[\begin{array}{cc}
m_{2} & 0\\
0 & Q_{c}
\end{array}\right].
\end{equation}

\noindent Note that following \citeA{Longuet-Higgins1957} notations, $m_{i}^{''}$ indicates the mixed wavenumber-frequency $i$-th spectral moment, where the number of quotes indicates the order of the frequency spectral moment, for example,

\begin{equation}
    m_3^{'} = \int_{0}^{\infty} 2 \pi f^{1}k^{3} E(k)dk, \label{eq:prime_moments}
\end{equation}

\noindent where $E(k)$ is a given wavenumber spectra.

Classically, to introduce $c$ in the joint density and obtain $p\left(c,\xi_{3}\right)_{0}$, we apply a change of variables

\begin{equation}
\xi_{2}=-\frac{\xi_{3}}{c}\;,\;\xi_{3}=\xi_{3}
\end{equation}

\noindent and after the integration of $p\left(c,\xi_{3}\right)_{0}$ over all the domain of definition of $\xi_{3}$, we obtain the distribution of $c$ (\citeA{Longuet-Higgins1957}, Eq. 2.5.19):

\begin{equation}
p\left(c\right)_{0}=\frac{1}{2}\frac{m_{4}m_{2}^{''}-m_{3}^{'}{}^{2}}{\sqrt{m_{4}}\left(c^{2}m_{4}+2cm_{3}^{'}+m_{2}^{''}\right)^{3/2}}
\end{equation}

\noindent Note that the sign on $c$ (or on ${\it m_{3}^{'}}$) depends on the convention on the wave propagation direction. We have kept the convention used by \citeA{Longuet-Higgins1957} here.

\subsubsection{Introducing the Orbital Velocity}

As indicated in Equation \ref{eq:Rice}, we can introduce in the formula a variable which represents the horizontal orbital velocity. For Gaussian waves the horizontal orbital velocity $u$ is defined as

\begin{equation}
u=\mathcal{H}_{t}\left(\frac{\partial\xi_{0}}{\partial t}\right)
\end{equation}

\noindent with $\mathcal{H}_{t}$ the Hilbert transform in time domain. Which means that

\begin{equation}
\xi_{0}=\sum_{i}a_{i}\cos\left(k_{i}x-\omega_{i}t\right)
\end{equation}

\noindent is transformed in

\begin{equation}
u=\sum_{i}a_{i}\omega_{i}\cos\left(k_{i}x-\omega_{i}t\right),
\end{equation}

\noindent with $a_i$ the wave amplitude, $k_i$ the wavenumber and $\omega_i$ the angular wave frequency of the wave component $i$. As the Hilbert transform is a linear operator, $u$ is also Gaussian. As previously, at the local maxima we have:

\begin{equation}
p\left(\xi_{2},\xi_{3},u\right)_{0}=\frac{\left.\left|\xi_{2}\right|p\left(\xi_{1},\xi_{2},\xi_{3},u\right)\right|_{\xi_{1}=0}}{N_{0}^{-}}
\end{equation}

\noindent with a new covariance matrix for $\xi_{1}$, $\xi_{2}$, $\xi_{3}$ and $u$:

\begin{equation}
Q=\left[\begin{array}{cccc}
m_{2} & 0 & 0 & 0\\
0 & m_{4} & m_{3}^{'} & m_{2}^{'}\\
0 & m_{3}^{'} & m_{2}^{''} & m_{1}^{''}\\
 & m_{2}^{'} & m_{1}^{''} & m_{0}^{''}
\end{array}\right]=\left[\begin{array}{cc}
m_{2} & 0\\
0 & Q_{c}
\end{array}\right].
\end{equation}

\noindent As previously, we can apply a similar change of variables

\begin{equation}
\xi_{2}=-\frac{\xi_{3}}{c}\;,\;\xi_{3}=\xi_{3}\;,\;u=u,
\end{equation}

\noindent or the easiest to deal with,

\begin{equation}
\xi_{3}=-c\xi_{2}\;,\;\xi_{2}=\xi_{2}\;,\;u=u
\end{equation}

\noindent  and integrate $p\left(c,\xi_{2},u\right)_{0}$ over all the domain of definition of $\xi_{2}$. The result is a  more complicated but again semi-analytical. The body of the integral has the form

\begin{equation}
{\rm e}^{-\frac{1}{2}\left[\xi\left(c\right)\xi_{2}^{2}+\beta(c,u)\xi_{2}+\alpha(u)\right]}\xi_{2}^{2}
\end{equation}

\noindent and its integration in $\xi_{2}$ on the down-crossings space $]-\infty,0]$ gives

\begin{equation}
I\left(c,u\right)=\frac{\left(\left(2\phi^{2}+1\right)\sqrt{\pi}\left({\rm erf}\left(\phi\right)+1\right){\rm e}^{\phi^{2}}+2\phi\right)}{\sqrt{2}\xi^{3/2}}{\rm e}^{-\alpha/2}
\end{equation}

\noindent  with

\begin{equation}
\phi=\phi\left(c,u\right)=\frac{1}{2\sqrt{2}}\frac{\beta(c,u)}{\sqrt{\xi\left(c\right)}},\quad\alpha=\alpha(u),
\end{equation}

\begin{equation}
\Delta=\det(Q_{c})=m_{3}^{'}\left(m_{2}^{'}m_{1}^{''}-m_{3}^{'}m_{0}^{''}\right)+m_{4}\left(m_{0}^{''}m_{2}^{''}-m_{1}^{''}{}^{2}\right)+m_{2}^{'}\left(m_{3}^{'}m_{1}^{''}-m_{2}^{'}m_{2}^{''}\right),
\end{equation}

\begin{equation}
\alpha(u)=\frac{m_{4}m_{2}^{''}-m_{3}^{'}{}^{2}}{\Delta}u^{2},
\end{equation}

\begin{equation}
\beta(c,u)=2\frac{m_{3}^{'}m_{1}^{''}-m_{2}^{'}m_{2}^{''}}{\Delta}u+2\frac{m_{4}m_{1}^{''}-m_{2}^{'}m_{3}^{'}}{\Delta}uc
\end{equation}

\noindent and

\begin{equation}
\xi\left(c\right)=\frac{m_{0}^{''}m_{2}^{''}-m_{1}^{''}{}^{2}}{\Delta}+2\frac{m_{3}^{'}m_{0}^{''}-m_{2}^{'}m_{1}^{''}}{\Delta}c+\frac{m_{4}m_{0}^{''}-m_{2}^{'}{}^{2}}{\Delta}c^{2}.
\end{equation}

The joint probability density of $\left(c,u\right)$ is then:

\begin{equation}
p\left(c,u\right)=\frac{1}{N_{0}^{-}}\frac{1}{(2\pi)^{2}\sqrt{m_{2}\Delta}}I\left(c,u\right)=\frac{I\left(c,u\right)}{2\pi\sqrt{m_{4}\Delta}}.
\label{eq:pcu}
\end{equation}

\noindent Note again that the sign on $c$ and $u$ (or on $m_{2}^{'}$ and $m_{3}^{'}$) depends on the convention on the wave propagation direction and \citeA{Longuet-Higgins1957}'s convention is still used here. The coefficients $\left(\alpha,\beta,\xi\right)$ can be calculated directly numerically and $\Delta$ is the determinant of $Q_{c}$, the sub-matrix of $Q$, and after the inverse of $Q_{c}$ is calculated:

\begin{equation}
Q_{c}^{-1}=\left[\begin{array}{cc}
R & \boldsymbol{s}\\
\boldsymbol{s}^{t} & r
\end{array}\right]
\end{equation}

\noindent  we find 

\begin{equation}
\alpha(u)=ru^{2}\label{eq:alpha},
\end{equation}

\begin{equation}
\beta(c,u)=2\left[\begin{array}{cc}
1 & c\end{array}\right]\boldsymbol{s}u,
\end{equation}

\begin{equation}
\xi\left(c\right)=\left[\begin{array}{cc}
1 & c\end{array}\right]R\left[\begin{array}{c}
1\\
c
\end{array}\right].
\end{equation}

\noindent An example of the joint density of the couple (phase speed, horizontal particle velocity) at local maxima in space is shown in  Figures \ref{fig:PhasevsVelocity}-a and b for a JONSWAP spectrum.

\subsection{Modelling $P_b$ from $p(c,u$)}\label{sec:pb_model}

By using Equation \ref{eq:pcu} applied to the dominant spectral wave band (that is, that contained in  the interval [$0.7f_p$, $1.3f_p$], where $f_p$ is the peak wave frequency), the probability of dominant wave breaking can be computed by integrating Equation \ref{eq:pcu} over all phase speeds and for orbital velocities over a threshold $A c$, with $A$ a constant that will be in the next section:

\begin{equation}
    P_b = \int_{u>Ac} \int_{0}^{\infty}p(c, u) dc du.
    \label{eq:Pbwork}
\end{equation}

\noindent $P_b$ will be modelled following Equation \ref{eq:Pbwork} hereafter. Note that from the definitions in Equation \ref{eq:N_0}, the proposed $P_b$ is defined as number of breaking local maxima over the total number of local maxima. From the analysis of $p(c,u)$ we observed that spurious, non-moving local maxima may exist around $c$ = $0$ and $u$ = $0$; therefore, to avoid artificially increasing $P_b$, we adopted a practical integration range of $c, u \in [0.05,+\infty]$ here. Note that this range may, however, only be valid for very narrow spectra. Further, we draw attention that, following from Equation \ref{eq:Rice}, our $P_b$ model is defined in space domain, whereas all the previous $P_b$ models and data are (at least partially) defined in time domain (see \ref{sec:review:parametric:models} for details). For the very narrow spectral band used here, the differences between temporal and spatial definitions of $Pb$ are negligible. This is discussed further in Section \ref{sec:discussion}.

Finally, the proposed model can be extended to accommodate two-dimensional spectra without changes on how $p(c,u)$ is calculated. This is done by applying an appropriated spreading function to any given one-dimensional spectra (or directly inputting a directional spectra) and by recalculating the moments in Equations \ref{eq:prime_moments} to take directionality into account or, more explicitly,

\begin{equation}
m_{i}=\int_{0}^{2\pi}\int_{0}^{\infty}\left( f \cos \theta \cos \alpha +  f \sin \theta \sin \alpha  \right)^{i} E(f,\theta)df d\theta.
\label{eq:mij}
\end{equation}

\begin{figure}[htp]
\includegraphics[width=0.95\linewidth]{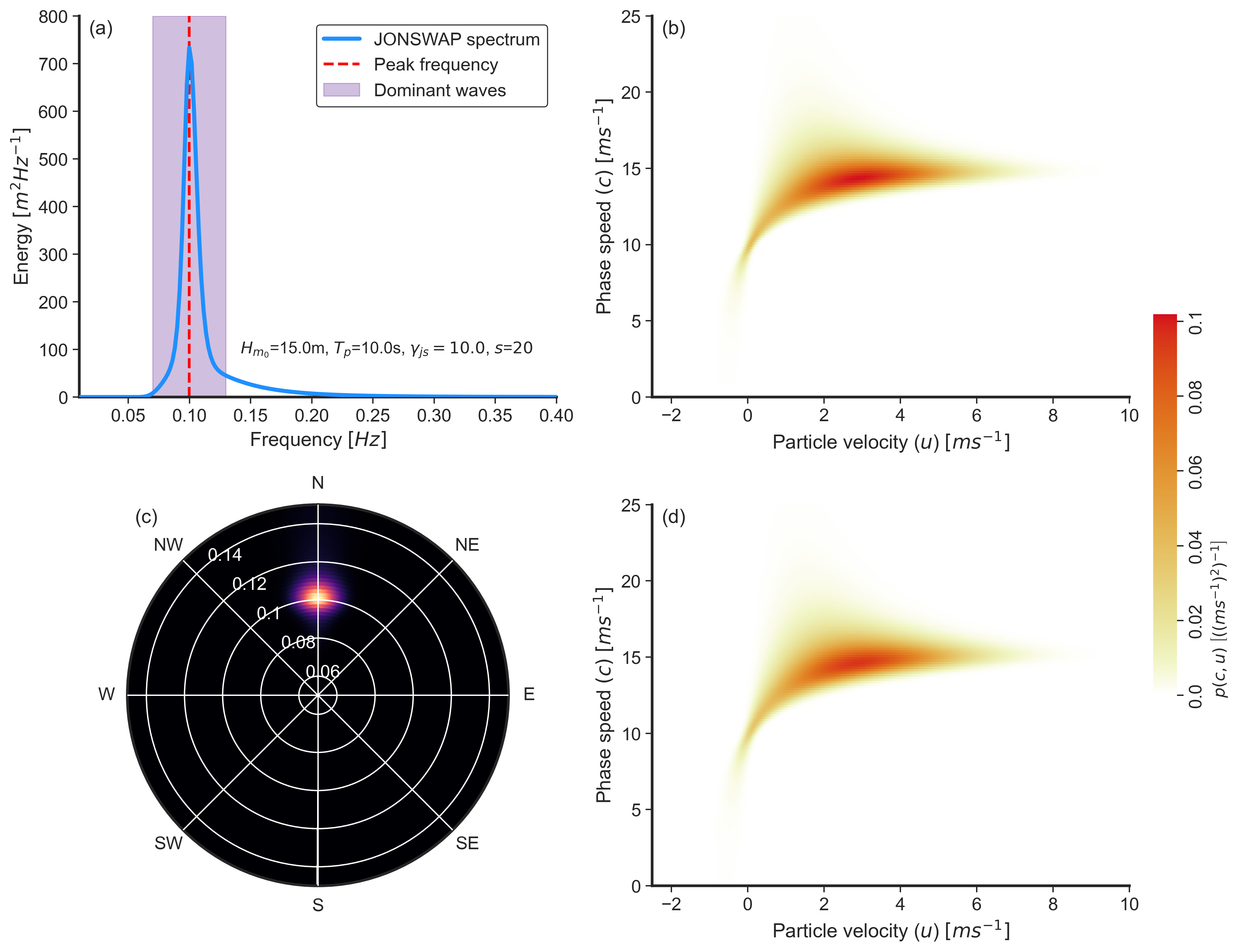}
\caption{Example of the application of the method. a) JONSWAP spectrum for $H_{m_0}$=15m, $T_p$=10s and shape parameter $\gamma_{js}$=10. b) Obtained joint probability density between the wave phase speed ($c$) and the horizontal particle velocity at wave crest ($u$) calculated using Equation \ref{eq:pcu}. Note that the joint probability density was computed using only the spectral energy between $0.7f_p$ and $1.3f_p$, that is, corresponding to the dominant wave band only. c) Directional spectrum for the same parameters as in a) and directional spreading $D(\theta)$ = $cos(\theta-\bar{\theta})^{2s}$ with $s$ = $20$ and $\bar{\theta}$ = $0$. d) Obtained $p(c, u)$ considering only the spectral energy in the direction $\alpha$ = $0$.  \label{fig:PhasevsVelocity}}
\end{figure}

\noindent An example considering the simplified cosine spreading law ($D(\theta)$ = $cos(\theta-\bar{\theta})^{2s}$) with $s$ = $20$, $\bar{\theta}$ = $0$ and $\alpha$=$0$ applied to same JONSWAP spectrum shown in Figure \ref{fig:PhasevsVelocity}-a is shown in Figure \ref{fig:PhasevsVelocity}-c. Note that the differences in $p(c,u)$ between the one-dimensional (Figure \ref{fig:PhasevsVelocity}-b) and the two-dimensional  (Figure \ref{fig:PhasevsVelocity}-d) spectra are negligible for the present assumptions. This relatively simple extension allows for the consideration of two-dimensional wave spectral but we caution the reader that it may not be fully complete. A follow-up publication will be dedicated to include and assess the effects of wave directionality in our method more rigorously.

\subsection{Definition of a Gaussian-equivalent Non-linear Wave Breaking Criterion}\label{sec:gaussian:waves:nonlinear}

The previously introduced joint probability density distribution $p(c,u)$ is based on Gaussian theory and therefore assumes that waves are linear. Breaking waves are, however, highly non-linear. For real non-linear waves, as detailed in the introduction, it is widely accepted that wave breaking starts when the water particle horizontal velocity at its crest ($u_{nl}$) reaches the wave phase speed ($c_{nl}$). A non-linear wave breaking criterion can be thus be defined as $A_{nl}$ = $u_{nl}/c_{nl}$ = $1$. Therefore, we assume that it is possible to obtain an equivalent kinematic criterion, $A_{lin}$ = $constant$ that relates Gaussian waves to non-linear waves.

Based on numerical experiments, \citeA{Cokelet1977} provided the potential and kinetic energy of a fully non-linear regular wave in deep-water at the onset of wave breaking (see the last row of his Table A.0). Based on his results, we define the kinematic criterion as the linear wave that has total energy equals to the nearly breaking non-linear regular wave computed by \citeA{Cokelet1977}. Following \citeA{Cokelet1977}, where $k$, $g$ and $\rho$ are expressed as non-dimensional variables, a deep-water wave at the breaking onset (see last row of his table A.0) has kinetic energy $T$ = $3.827 \times 10^{-2}$ and potential energy $V$ = $3.457 \times 10^{-2}$. The energy-equivalent linear wave (denote with subscript $eq$) has, therefore, amplitude:

\begin{equation}
a_{eq} = \sqrt{2 \times E} = \sqrt{2 \times (V+T)} = 0.3817.
\end{equation}

\noindent For this particular case, the linear dispersion relation reads:

\begin{equation}
\omega^2=gk=1, 
\end{equation}

\noindent  the fluid velocity at crest of the energy-equivalent linear wave is:

\begin{equation}
u_{eq} = \omega a_{eq} = 0.3817,
\end{equation}

\noindent and the phase speed of the linear wave is:

\begin{equation}
c_{eq} = \sqrt{\frac{g}{k}} = 1.
\end{equation}

\noindent Given these constants, we obtain:

\begin{equation}
A_{lin} = \frac{u_{eq}}{c_{eq}} = \frac{0.3817}{1} = 0.3817.
\end{equation}

Following this approach, we define the correction coefficient $A$ = $A_{lin}$ = $0.382$ that will be used as reference value hereafter for our tests. This result is consistent with recent findings from \citeA{Ardag2020} who reported from the re-analysis of \citeauthor{Duncan1981}'s \citeyear{Duncan1981} experimental results, a wave breaking threshold between 0.75 and 1.02 (see their Figure 1). Note, however, that these authors defined their wave breaking threshold as $u/c_g$, where $c_g$ is the group velocity and $u$ was obtained from linear wave theory. Replacing wave group velocity ($c_g$) by the wave phase speed ($c$) yields a range of possible values between 0.35 and 0.50, which is consistent with $A_{lin}$.

\begin{figure}[htp]
	\includegraphics[width=0.99\linewidth]{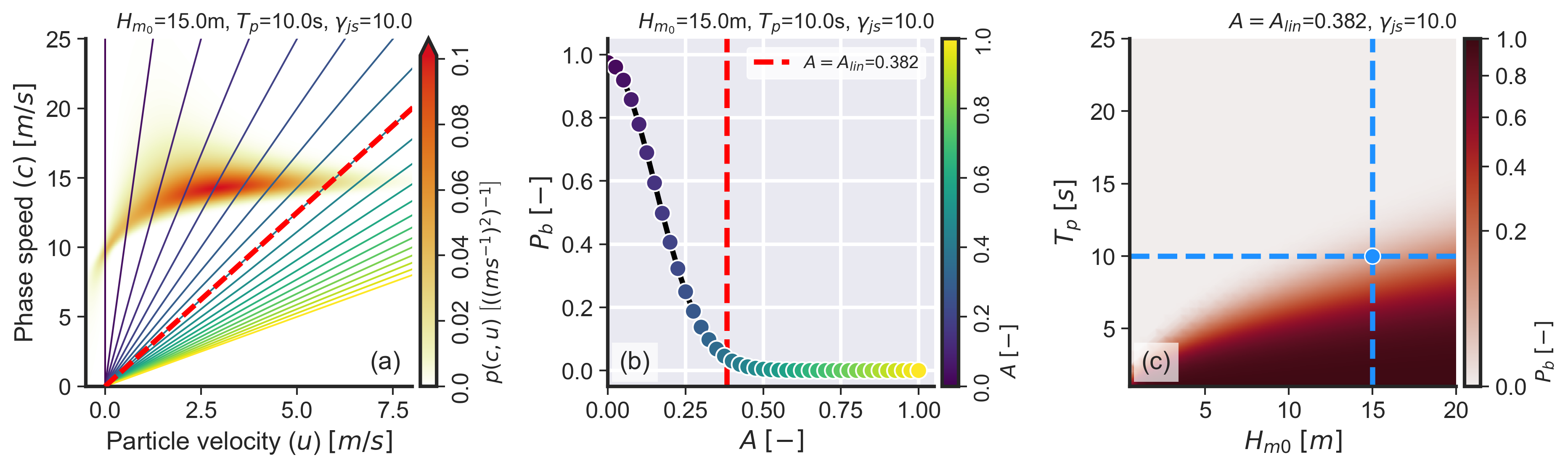}
\caption{a) Example of joint probability density between $u$ and $c$ obtained from Equation \ref{eq:Pbwork}. The colored lines indicate different values of $A$ and the red dashed line shows $A$=$A_{lin}$=0.382. b) Possible values of $P_b$ for varying $A$ calculated using the joint PDF from a). The vertical dashed line shows $A$=$A_{lin}$=0.382. c) Obtained $P_b$ for varying $H_{m_0}$ and $T_p$ and fixed $A$ (0.382) and $\gamma_{js}$ (10). The dashed blue lines and marker indicate the $H_{m_0}$ and $T_p$ values used in a) and b). Note that as in Figure \ref{fig:PhasevsVelocity}, these results only consider dominant waves, that is, they were calculated from the spectrum between $0.7f_p$ and $1.3f_p$.}\label{fig:sensitivity}
\end{figure}

Figure \ref{fig:sensitivity} illustrates the sensitivity in wave breaking probability with changes in the wave breaking threshold $A$. For the given $p(c, u)$ in Figure \ref{fig:sensitivity}-a, letting $A$ to vary from $0$ to $1$ resulted in a exponential increase in $P_b$ at $A\le0.2$ (Figure \ref{fig:sensitivity}-b), which may be unrealistic. When setting $A$=$A_{lin}$=0.382 and letting the significant wave height ($H_{m_0}$) and wave peak period ($T_p$) vary in the definition of the JONSWAP spectrum, the results indicate that steeper waves are more probable to break, which is expected (Figure \ref{fig:sensitivity}-c). Finally, note that the wave breaking threshold $A$ might be sensitive to other wave and atmospheric parameters such as wave directionality or direct wind forcing (or, equivalently, wave age). In the next sections, the accuracy of our model is assessed using field observations and our results are compared with other parametric wave breaking formulations.

\section{Field Data}\label{sec:materials:methods:field:data}

Three historical datasets were used to evaluate the present model. Further, six historical models (detailed in \ref{sec:review:parametric:models}) were chosen to contextualize our model in relation to the state-of-the-art. These historical models range from baseline models in which the only inputs are known environmental parameters (wind speed in \citeA{Melville2002} or wave steepness in \citeA{Banner2000}, for example) to fairly complex models that account for combinations of several phenomena (\citeA{Romero2019}, for example).

\subsection{Thomson (2012) and Schwendeman et al. (2014) dataset (TSG14)}\label{sec:field:data:thomson:schwendeman}

The first data are from \citeA{Thomson2012} and \citeA{Schwendeman2014}, hereafter TSG14, and were collected in the Strait of Juan de Fuca, Washington. These data were collected by a gray scale video camera with a resolution of $640 \times 480$ pixels installed above the wheelhouse of Research Vessel R/V \textit{Robertson} which recorded at an acquisition rate of $30$ Hz \cite{Schwendeman2014}. These data were then projected into a metric coordinate grid with resolution of 0.25m (cross wave) and 0.075m (along wave) using the method proposed by \citeA{Holland1997} and were then used to obtain $\Lambda(c)$ using the spectral approach of \citeA{Thomson2009a}. The data were collected in a (usually) fetch-limited region and for a young sea state; note, however, that the particular sea-states analyzed here may not be fetch-limited. Figure \ref{fig:data:plot}-a shows the measured wave spectra, Figure \ref{fig:data:plot}-b shows $\Lambda(c)$ distributions, and Table \ref{tab:data:summary} shows a summary of these data. For these data, $P_b$ was calculated using the measured $\Lambda(c)$ distributions combined with the method described below in Equation \ref{eq:pb:from:lambda}. Additional information regarding the data collection is available from \citeA{Thomson2012} and \citeA{Schwendeman2014}.

\subsection{Sutherland and Melville (2013) dataset (SM13)}\label{sec:field:data:sutherland}

\begin{figure}[htp]
	\includegraphics[width=0.99\linewidth]{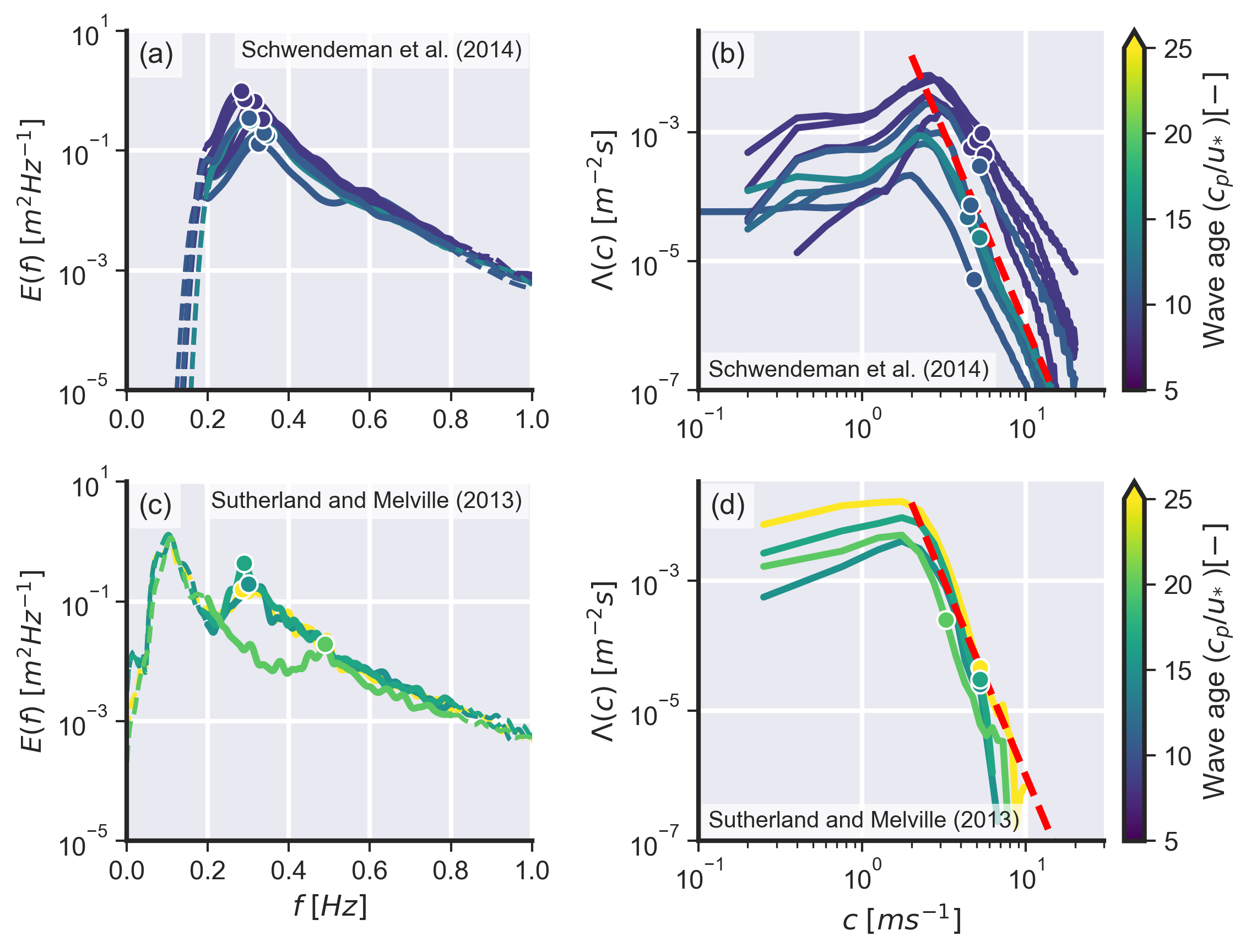}
	\caption{Field data. a) Spectral data from TSG14. b) $\Lambda(c)$ data from STG14. c) Spectral data from SM13. c) $\Lambda(c)$ data from SM13. The coloured circular markers show in a) and c) show the peak frequency ($f_p$) and the coloured circular markers show in b) and d) show the peak wave speed ($c_p$). The red dashed line in b) and d) shows the theoretical $c^{-6}$ decay predicted by \citeA{Phillips1985}. In all plots, the color scale shows the wave age ($c_p/u_{*}$). }\label{fig:data:plot}
\end{figure}

The second dataset is from \citeA{Sutherland2013}, hereafter SM13, and was collected using the Research Platform R/P \textit{FLIP} during a two-day field campaign in the Southern California Bight under the scope of the SoCal 2010 experiment \cite{Sutherland2013}. Here, we focus only on the visible imagery collected by these authors to keep consistency with the previously presented data. Stereo video data were collected by a pair of video cameras mounted on the R/P FLIP for 10 minutes at the start of each hour and $\Lambda(c)$ was obtained using a variation of the method of \citeA{Kleiss2011}, that is, tracking the temporal evolution of breakers obtained via pixel intensity threshold. Figure \ref{fig:data:plot}-c shows the measured wave spectra, Figure \ref{fig:data:plot}-d shows $\Lambda(c)$ distributions, and Table \ref{tab:data:summary} shows a summary of these data. Note that because wave breaking was not observed for frequencies below 0.2$Hz$ and from numerical simulations (not shown) these waves corresponded to a cross-swell not forced by the wind, our analyses only consider waves in the frequency range $0.2 <f < 0.8Hz$.  Additional information regarding the data collection is available from \citeA{Sutherland2013}. For these and TSG14 data, $P_b$ was calculated using the measured $\Lambda(c)$ distributions combined with the formulas from \citeA{Banner2010}:

\begin{equation}
    P_b = \frac{\int_{c_0}^{c_1} c\Lambda(c)dc}{\int_{c_0}^{c_1} c\Pi(c)dc}\label{eq:pb:from:lambda}
\end{equation}

\noindent where $c_0=\frac{g}{2\pi}\frac{1}{1.3f_{p}}$, $c_1=\frac{g}{2\pi}\frac{1}{0.7f_{p}}$, $\Pi(c)= \chi g / (2 \pi c^3)$ and $\chi=0.6$. The implication of this choice is discussed in further detail in Section \ref{sec:discussion}.

\subsection{Banner, Babanin and Young (2000) dataset (B00)}\label{sec:field:data:banner}

The third dataset is from \citeA{Banner2000}, hereafter B00, and was collected in the Black Sea (BS), Lake Washington (LW) and the Southern Ocean (SO). These authors directly provide values for significant wave height $H_{m0}$, peak period ($T_p$) and the wave breaking probability in their Tables 1 (Black Sea, denoted as BS here) and 2 (Southern Ocean, denoted as SO here). The majority of the data were collected in the Black Sea (13 data runs) and two data runs are from the Southern Ocean. Given that the original spectral data were not published alongside their paper, we approximate the observed spectra using the provided pairs $H_{m0}$, $T_p$ assuming a JONSWAP shape with $\gamma_{js}=3.3$ (as previously done in \citeA{Filipot2010}, for example). Given that in this paper we are only interested in a very narrow spectral band, the differences between observed and simulated spectra should be minimal. For more details regarding this data refer to \citeA{Banner2000}.

\begin{table}[htbp]
  \centering
  \caption{Data summary for the two experiments described in Sections \ref{sec:field:data:thomson:schwendeman} and \ref{sec:field:data:sutherland}. Note that the parameters obtained from wave spectra were computed specifically for the bands shown in Figure \ref{fig:data:plot} for TSG14 and SM13 cases. The wave height ($H_p$) and wave steepness ($\epsilon$) parameters for dominant waves were calculated as per \citeA{Banner2002} (see Section \ref{sec:parametric:models:banner:2000} for details). The wave age parameter was calculated as $c_p/u_*$ .}
  \resizebox{.99\textwidth}{!}{%
    \begin{tabular}{llllllllllll}
    \toprule
    Dataset & Date  & Length & $H_{m_0}$ & $T_p$ & $H_p$ & $\epsilon$ & $U_{10}$ & $u_*$ & $c_p$ & Wave age & $P_b$ \\
    \hline
     & $[-]$  & $[min]$ & $[m]$ & $[s]$ & $[m]$ & $[-]$ & $[ms^{-1}]$ & $[ms^{-1}]$ & $[ms^{-1}]$ & $[-]$ & $[-]$ \\
    \hline
    TSG14 & 14/02/2011 20:33 & 6.5 & 0.75  & 2.88  & 0.66  & 0.160 & 11.50 & 0.373 & 4.50  & 12.07 & 3.54E-03 \\
    TSG14 & 14/02/2011 20:58 & 5.1 & 0.75  & 2.96  & 0.66  & 0.152 & 12.55 & 0.417 & 4.62  & 11.08 & 9.57E-03 \\
    TSG14 & 14/02/2011 21:30 & 6.5 & 0.91  & 2.99  & 0.82  & 0.184 & 15.07 & 0.561 & 4.67  & 8.33  & 6.29E-02 \\
    TSG14 & 14/02/2011 21:44 & 8.5 & 1.09  & 3.17  & 1.00  & 0.200 & 15.73 & 0.599 & 4.94  & 8.25  & 1.01E-01 \\
    TSG14 & 14/02/2011 22:29 & 6 & 1.21  & 3.44  & 1.09  & 0.186 & 17.24 & 0.636 & 5.36  & 8.44  & 1.51E-01 \\
    TSG14 & 14/02/2011 22:37 & 4.8 & 1.37  & 3.53  & 1.24  & 0.199 & 18.01 & 0.660 & 5.52  & 8.36  & 7.61E-02 \\
    TSG14 & 15/02/2011 19:04 & 10 & 0.87  & 3.29  & 0.79  & 0.146 & 14.45 & 0.360 & 5.13  & 14.28 & 3.75E-03 \\
    TSG14 & 15/02/2011 19:19 & 6 & 0.90  & 3.31  & 0.81  & 0.149 & 13.11 & 0.477 & 5.17  & 10.85 & 4.05E-02 \\
    SM13  & 06/12/2010 21:59 & 10 & 0.61  & 3.51  & 0.52  & 0.085 & 6.46  & 0.205 & 5.48  & 26.68 & 7.96E-03 \\
    SM13  & 06/12/2010 23:00 & 10 & 0.61  & 3.33  & 0.54  & 0.097 & 7.55  & 0.342 & 5.20  & 15.22 & 1.95E-03 \\
    SM13  & 07/12/2010 00:00 & 10 & 0.73  & 3.45  & 0.66  & 0.112 & 8.62  & 0.319 & 5.38  & 16.85 & 3.24E-03 \\
    SM13  & 08/12/2010 00:00 & 10 & 0.34  & 2.04  & 0.23  & 0.110 & 5.24  & 0.160 & 3.19  & 19.96 & 1.65E-02 \\
    B00 (SO) & 10/6/1992 & 5 & 9.20  & 13.46 & 8.02  & 0.089 & 19.80 & 0.835 & 21.01 & 25.17 & 2.70E-02 \\
    B00 (SO) & 11/6/1992 & 9 & 4.20  & 12.04 & 3.66  & 0.051 & 16.00 & 0.626 & 18.78 & 30.02 & 0.00E+00 \\
    B00 (BS) & 1993  & 34-68 & 0.39  & 2.78  & 0.34  & 0.089 & 11.70 & 0.414 & 4.34  & 10.49 & 3.80E-02 \\
    B00 (BS) & 1993  & 34-68 & 0.49  & 2.94  & 0.43  & 0.100 & 12.70 & 0.461 & 4.59  & 9.96  & 6.50E-02 \\
    B00(BS) & 1993  & 34-68 & 0.53  & 3.33  & 0.47  & 0.084 & 14.00 & 0.524 & 5.20  & 9.93  & 6.00E-02 \\
    B00 (BS) & 1993  & 34-68 & 0.54  & 3.23  & 0.47  & 0.092 & 14.40 & 0.544 & 5.04  & 9.26  & 5.20E-02 \\
    B00 (BS) & 1993  & 34-68 & 0.38  & 2.27  & 0.34  & 0.131 & 15.00 & 0.574 & 3.55  & 6.18  & 6.30E-02 \\
    B00 (BS) & 1993  & 34-68 & 0.45  & 2.56  & 0.40  & 0.121 & 14.60 & 0.554 & 4.00  & 7.23  & 6.70E-02 \\
    B00 (BS) & 1993  & 34-68 & 0.45  & 2.44  & 0.40  & 0.134 & 13.70 & 0.509 & 3.81  & 7.49  & 8.40E-02 \\
    B00 (BS) & 1993  & 34-68 & 1.19  & 5.88  & 1.04  & 0.061 & 8.70  & 0.295 & 9.18  & 31.10 & 0.00E+00 \\
    B00 (BS) & 1993  & 34-68 & 1.32  & 6.24  & 1.15  & 0.060 & 11.20 & 0.391 & 9.74  & 24.91 & 0.00E+00 \\
    B00 (BS) & 1993  & 34-68 & 0.83  & 6.24  & 0.73  & 0.038 & 9.50  & 0.322 & 9.74  & 30.22 & 0.00E+00 \\
    B00 (BS) & 1993  & 34-68 & 0.89  & 5.88  & 0.78  & 0.045 & 10.70 & 0.368 & 9.18  & 24.91 & 0.00E+00 \\
    B00 (BS) & 1993  & 34-68 & 0.99  & 3.71  & 0.87  & 0.127 & 10.00 & 0.339 & 5.79  & 17.06 & 3.40E-02 \\
    B00 (BS) & 1993  & 34-68 & 0.88  & 4.00  & 0.77  & 0.097 & 8.70  & 0.295 & 6.24  & 21.14 & 5.80E-02 \\
    \bottomrule
    \end{tabular}%
    }
  \label{tab:data:summary}%
\end{table}%

\section{Results}\label{sec:results}

\subsection{Comparison with Field  Data}\label{sec:results:comparison}

\begin{figure}[htp]
	\includegraphics[width=0.99\linewidth]{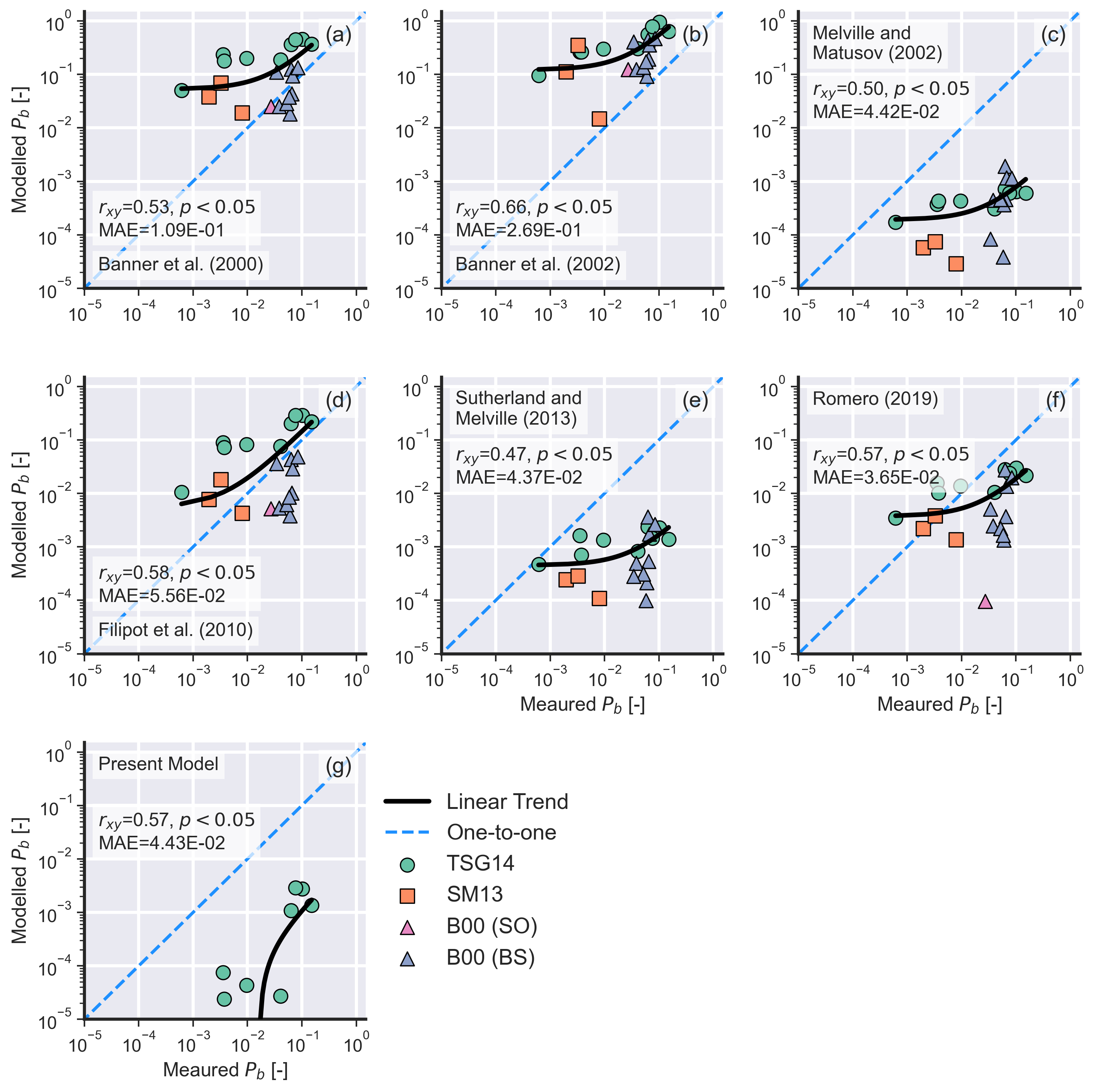}
	\caption{Compassion between measured and computed $P_b$ for different models and data. a) \citeA{Banner2000}, b) \citeA{Banner2002}, c) \citeA{Melville2002}, d) \citeA{Filipot2010}, e) \citeA{Sutherland2013}, f) \citeA{Romero2019}, and g) present model with $A$=$A_{lin}$=0.382. The thick black line shows the linear regression between measured and modelled $P_b$ and the blue dashed line indicates the one-to-one correspondence in all panels. Data points with modelled $P_b$ $<$ $10^{-5}$ or observed $P_b$ = $0$ are not shown in this plot. In all plots, $r_{xy}$ is Pearson's correlation coefficient and $MAE$ indicates the mean absolute error. Note the logarithmic scale.}\label{fig:one_to_one_comparison}
\end{figure}

Figure \ref{fig:one_to_one_comparison} shows the comparison between estimated (or observed) (x-axis) and modelled (y-axis) values of $P_b$ for each model. In general, no  model was able to closely reproduce the trends seen in the combined observed data, regardless of the underlying mathematical or physical formalism. Furthermore, orders of magnitude of difference between the models and, more worryingly, between the models and the measured data were observed. In general, models based on a wave steepness-derived wave breaking criterion (\citeA{Banner2000}, \citeA{Banner2002}, for example) overestimated data derived from $\Lambda(c)$ while models based on $\Lambda(c)$ (\citeA{Melville2002} and \citeA{Sutherland2013}, for example) underestimated $P_b$ data that was not derived from $\Lambda$ (that is, B00 data). The model from \citeA{Filipot2010} was found to be the most consistent model. From Figure \ref{fig:one_to_one_comparison}-g, the formulation presented in Section \ref{sec:gaussian:waves} with $A$ = $A_{lin}$ = 0.382 underestimated the observed $P_b$ for B00 and SM13 data (note that $P_b$ was too low to be displayed on the plot) but performed relatively well for the majority of TSG14 data. Using the mean absolute error (MAE) as a convenient metric to assess the models, it was found that the present model has errors in the same order of magnitude as the previous models. Given the spread in the results seen in Figure \ref{fig:one_to_one_comparison}, no model could be considered a clear winner. For the discussion of these results, see Section \ref{sec:discussion}.

\subsection{Model Optimization}\label{sec:results:optmization}

\begin{figure}[htp]
	\includegraphics[width=0.95\linewidth]{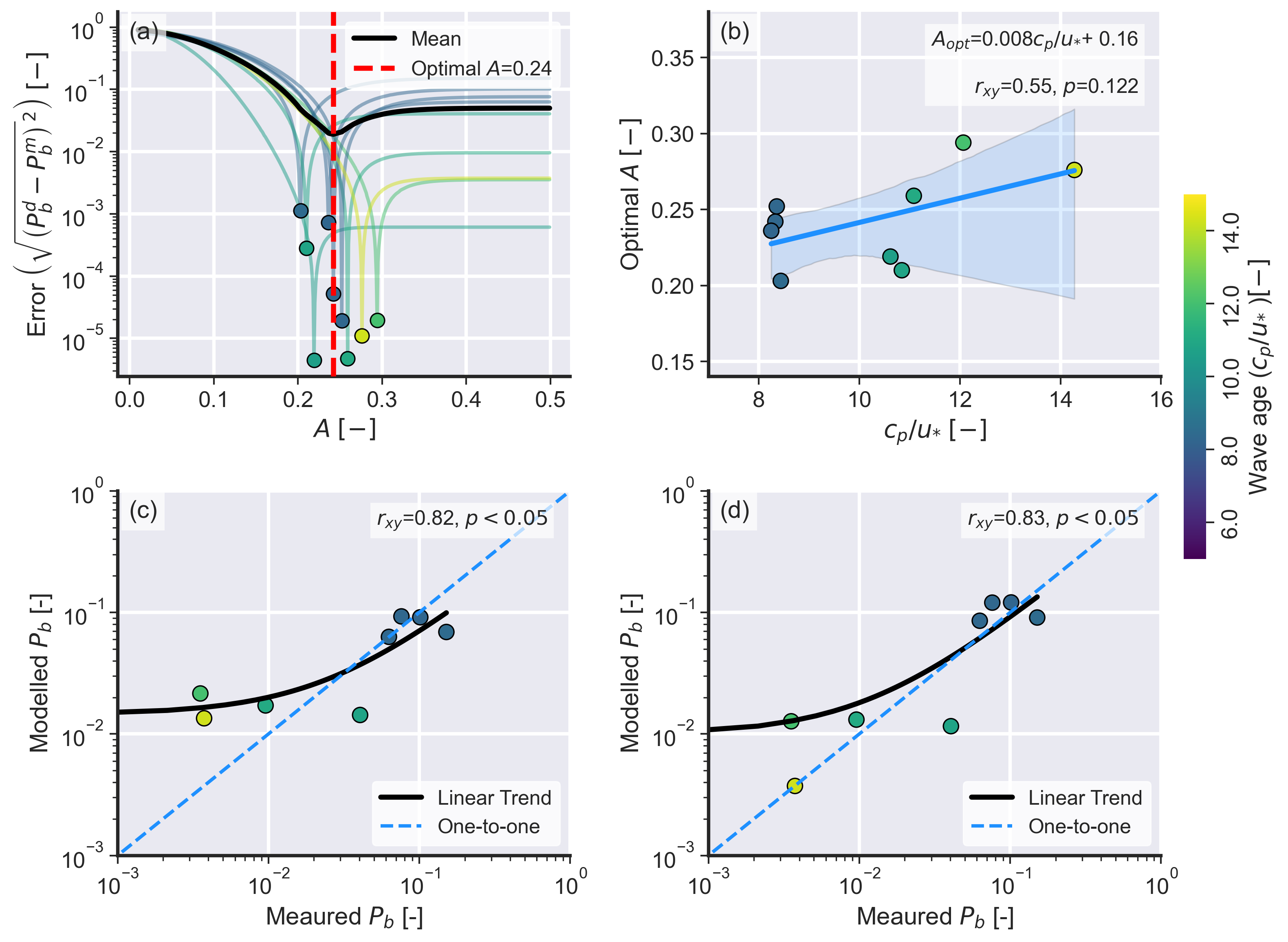}
	\caption{Results of the optimization procedures. a) Optimization curves for each data record (coloured lines) and the global averaged  (black line). The vertical dashed line show s$A$ = $A_{opt}$ = 0.24. b) Parametrization of $A$ as a function of $c_p/u_*$. The blue swath indicates the 95\% confidence interval. For this particular case, $A$ = $0.008 c_p/u_*$ $+$ $0.16$. Note the logarithmic scale in a), c) and d). In all plots, the color scale shows the wave age ($c_p/u_{*}$). In b) to d) $r_{xy}$ is Pearson's correlation coefficient.}\label{fig:optimization}
\end{figure}

From the analysis of Figure \ref{fig:sensitivity}, minor changes in $A$ can lead to major variations in $P_b$. Further, from the analysis of Figure \ref{fig:one_to_one_comparison}, the proposed model underestimated $P_b$ for $A$ = $A_{lin}$ = 0.382 particularly for S13 and B00 data. Given that it is a common practice to optimize wave breaking models for particular datasets, we present two methods to do so using TSG14 data as an example. The same could be done for B00 and SM13 data but, for brevity, this is not done here. Given that the present model is not computationally expensive, the first approach consisted of varying $A$ from 0.1 to 0.5 in 0.001 intervals and finding the value of $A$  that resulted in the lowest squared error ($\sqrt{\left( p_{b_{i}}^d - p_{b_{i}}^m \right)^2}$, where the superscripts $d$ and $m$ indicate observed and modelled data, respectively) for each data run. Figure \ref{fig:optimization}-a shows the results of this procedure. The value $A$ = $A_{opt}$ = 0.24 was, on average, the optimal values of for this particular dataset. The second approach consisted in parameterizing the optimal value of $A$ for each data run as a function of a known environmental variable, in this example, the waveage $c_p/u_*$ (Figure \ref{fig:optimization}-b). The results of these two approaches are show in Figures \ref{fig:optimization}-c and d, respectively. Both approaches considerably improved the model results from the baseline model presented in Figure \ref{fig:one_to_one_comparison}, with the parametric model (Figure \ref{fig:optimization}-d) performing slightly better when considering Pearson's correlation coefficient ($r_{xy}$) as a comparison metric.

\section{Discussion}\label{sec:discussion}

We have introduced a new model for obtaining the probability of wave breaking ($P_b$) for dominant waves based on the theoretical joint probability density distribution between wave phase speed ($c$) and horizontal orbital velocity at the wave crest ($u$) for unidirectional Gaussian wave fields. The present model has only one parameter for defining the wave breaking threshold ($A$), which makes it relatively easy to optimize for a given dataset (as shown in Section \ref{sec:results:optmization}). While the proposed model performed relatively well for one of the investigated datasets (TSG14), it greatly underestimated $P_b$ for the two other datasets (SM13 and B00). For the data investigated here, such underestimation did not result in a high mean absolute error (MAE) and, in fact, our model had one of the lowest MAE. Recent results of \citeA{Barthelemy2018}, \citeA{Derakhti2020} and \citeA{Varing2020} showed that waves with horizontal fluid velocity that exceeds 0.85 times the phase velocity will inevitably break. These results suggest that the breaking threshold derived from \citeA{Cokelet1977} in Section \ref{sec:gaussian:waves:nonlinear} could be reduced by $\approx$15\%. If we apply their findings to our case, we obtain $A$ $=$ $0.382$ $\times$ $0.85$ $=$ $0.324$ which would help to reduce the underestimation of $P_b$, but not significantly. It is more probable that other environmental phenomena such as direct wind forcing, directional spreading and long wave modulation, which are not accounted in our model, are the reason for such differences. 

One of the most challenging aspects when assessing our model is, nevertheless, regarding the field data. The attribution of wave breaking occurrences to wave scales using timeseries analysis, as done in \citeA{Banner2000} or \citeA{Filipot2010}, is difficult because several wave scales can be present at the same time and space. This lead us to use $\Lambda(c)$ observations as well as data from \citeA{Banner2000} to investigate our model. Different interpretations of how $\Lambda(c)dc$ is computed from field data can, however, generate orders of magnitude of difference in its moments \cite{Gemmrich2013, Banner2014} and, consequently, in $P_b$. Next, it is difficult to relate the speed of the wave breaking front to the phase speed of the carrying wave because small, slower breaking waves could merely be traveling on top of longer, much faster waves. In particular, we believe that these wave breaking events can significantly contribute to the observed $\Lambda(c)dc$ distribution as they would have $c$ close to the peak wave phase speed. This wave breaking ``sub-population'' has not receive much research interest because of its apparent small contribution to energy dissipation but, for our particular case, they directly impact model validation.

Further, relating $\Lambda(c)$ to $P_b$ is also challenging. Here, we adopted the convenient formula from \citeA{Banner2010}. While this formula has some support from the literature \cite{Ardhuin2010}, the actual functional form of $\Pi(c)$ and the value for the constant $\chi$ (see Equation \ref{eq:pb:from:lambda}) are unknown and changes in these will lead to changes in $P_b$. The Gaussian framework developed in Section \ref{sec:gaussian:waves:1} provides an alternative method to obtain $\Pi(c)$ (from Equation \ref{eq:N_0}, for example) but this is beyond the scope of this introductory paper and will be the focus of a future publication.

Finally, we would like to re-emphasize that our model is derived in the space domain whereas $P_b$ data is (at least partially) obtained in the time domain. For the narrow spectral band investigated here, Monte-Carlo simulations of linear waves indicate that the difference between $P_b$ modelled in space is less than five percent from $P_b$ modelled in time (not shown). Given all these complications and the fact that some historical models are being compared to data that was used to create them (\citeA{Banner2000} and \citeA{Sutherland2013}, for example), we are unable to provide an accurate ranking of the existing models. Future research should focus, therefore, on obtaining $P_b$ data that is unambiguous and widely available. In this regard, and despite its own limitations, wave tank experiments could bring further insight on the statistics of dominant (or not) breaking waves. Such a dataset would ultimately allow researchers to focus on models derived from physical and mathematical concepts (such as ours) rather than on empirical concepts. 

\section{Conclusion}\label{sec:conclusion}

We have presented a new statistical wave breaking model derived from Gaussian field theory that we have applied to obtain the probability of wave breaking for dominant, wind-sea waves. Although more mathematically complex than previous formulations, the present model relies on the ratio between the crest orbital velocity and the phase speed and uses only on a single free parameter, the wave breaking threshold $A$. Using theoretical results obtained by \citeA{Cokelet1977} for regular nearly breaking waves, we derived a wave breaking threshold to adapt our linear model to non-linear waves. The present model has errors in the same order of magnitude as six other historical models when assessed using three field datasets. For a particular dataset (TSG14), our model performed well, especially if the free-parameter $A$ is fine tuned. Additional observations are however required, to further understanding and quantifying the dependence of $A$ on environmental parameters that are not accounted for in our model (for example, wind forcing, wave directionality or modulation by long waves). Future research should be dedicated to collect more wave breaking observations in different and repeatable environmental conditions to provide reliable constraints for the optimization of the present and other wave breaking models. Still and although the research presented here is in early stages, the present model should be extendable to waves of any scale and, therefore, has the potential to be implemented in current state-of-the-art spectral wave models as a new wave breaking dissipation source term with relatively little effort.

\appendix

\section{Historic Parametric Wave Breaking Models}\label{sec:review:parametric:models}

\subsection{Banner et al. (2000)}\label{sec:parametric:models:banner:2000}

\citeauthor{Banner2000}'s \citeyear{Banner2000} is a popular model for calculating wave breaking probabilities for deep water, dominant waves. 
This model follows from observations and results from 
 \citeA{Donelan&al.1972}, \citeA{Holthuijsen&Herbers1986} and
 \citeA{Banner&Tian1998} who demonstrated the importance of the wave group modulation on the wave breaking onset. These authors conveniently obtained a parameterization for the probability of wave breaking ($P_b$) based solely on the spectral steepness of the dominant wave scale ($\epsilon_p$), assuming that their formulas would capture the influence of the wave group modulation on the wave breaking onset. Their formulation was derived using a dataset of measurements collected in various environments ranging from lakes to open ocean conditions \cite{Banner2000}. From these observations, these authors were then able to obtain a wave breaking threshold behaviour for the dominant waves as a function of the dominant spectral wave steepness given by:

\begin{equation}
    \epsilon_p = \frac{H_p k_p}{2} \label{eq:02:04}
\end{equation}

\noindent in which $k_p$ is the wavenumber at peak frequency ($f_p$) and $H_p$ is the significant wave height of the dominant waves calculated as:

\begin{equation}
    H_p = 4 \sqrt{\left( \int_{0.7f_p}^{1.3f_p} E(f)df\right)} \label{eq:02:05}
\end{equation}

\noindent where $E(f)$ is the spectra of wave heights as a function of frequency. For their data, $P_b$ was then parameterized as a single equation with three free parameters ($p_1$, $p_2$, $p_3$):

\begin{equation}
	P_b = p_1+(\epsilon_p-p_2)^{p_3}, \label{eq:02:03}
\end{equation}

\noindent For the available field data, \citeA{Banner2000} found optimal values of $p_1=22$, $p_2=0.055$, and $p_3=2.01$. Note that hereafter free parameters for the different models will be denoted as $p_n$ where $n$ is a sequential number.

\subsection{Banner et al. (2002)}\label{sec:parametric:models:banner:2002}

This work extended \citeA{Banner2000} model to shorter wave scales (up to 2.48 times the peak wave frequency). From field data  \citeA{Banner2002} reported that the waves were breaking if the saturation spectrum $\sigma(f) = 2\pi^{4}f^{5} E(f)/2 g^2 = \sigma(k) = k^4E(k)$  exceeded a threshold that was frequency dependent. These author's related this dependence to the directional spreading $\overline{\theta(k)}$ which later led \citeA{Banner2010} to explicitly define the following empirical formulation:

\begin{equation}
P_b(k_c)= \mathcal{H}_h(\tilde{\sigma}(k_c)-p_1) \times p_2 \times (\tilde{\sigma}(k_c)-\tilde{\sigma}_{t}),
\label{Q_BGF}
\end{equation}

\noindent in which $\mathcal{H}_h$ is the Heaviside step function, $k_c$ is the central wavenumber for a given wavenumber range, $\tilde{\sigma}(k_c)=\sigma(k_c) / \overline{\theta(k_c)}$ is the saturation spectrum normalized by the averaged directional spreading, $p_1=0.0045$ and $p_2$ = $33$ are constants obtained from their observations. Following \citeA{Banner2002}, the directional spreading angle is calculated according to \citeA{Hwang2000} (their equation 19a):

\begin{equation}
    \theta\left(\frac{k}{k_p}\right) =
\left\{
	\begin{array}{ll}
		0.35 + 1.05\left(1-\frac{k}{k_p}\right)  & \mbox{if } \frac{k}{k_p} < 1.05 \\
		0.30 + 0.087\left(\frac{k}{k_p}-1\right)  & \mbox{if } 1.05 \leq \frac{k}{k_p} < 5 \\
	\end{array}
\right.
\end{equation}

\noindent where $\theta$ is the directional spreading angle as a function of the wavenumber.

\subsection{Filipot et al. (2010)}\label{sec:parametric:models:filipot:2010}

This method follows from the original works of \citeA{LeMehaute1962}, \citeA{Battjes1978} and \citeA{Thornton1983} and assumes that the probability distribution function (PDF) of breaking wave heights in the dominant wave scale is parameterized by its central frequency $f_c$ or, equivalently, by its representative phase speed $c(f_c)$ and the product between a Rayleigh PDF for the wave heights

\begin{equation}
	P(H, f_c) = \frac{2H}{H_{rms}^{2}(f_c)} \exp{\left[- \left( \frac{H}{H_{rms}(f_c)} \right)^2 \right] } \label{eq:02:02:01}
\end{equation}

\noindent in which

\begin{equation}
	H_r(f_c) = \frac{4}{\sqrt{2}} \sqrt{\int_{0}^{\infty} U_{fc}(f)E(f)df } \label{eq:02:02:02}
\end{equation}

\noindent and

\begin{equation}
	U_{f_c} = 0.5 - 0.5 \cos \left( \frac{\pi}{\delta} \left[ \frac{f}{f_c} - 1 - \delta \right] \right) \label{eq:02:02:03}
\end{equation}

\noindent where $\delta$ is the bandwidth of a Hann window (in this study, $\delta=0.6$), and a weighting function

\begin{equation}
	W(H, f_c) = p_1\left[\frac{\beta_r}{\beta}\right]^2 \left\{1 - \exp{\left[-\left(\frac{\beta}{\tilde{\beta}}\right)^{p_2} \right] }  \right\} \label{eq:02:02:04}
\end{equation}

\noindent in which $\beta=kH/\tanh(kh)$, and $p_1$ and $p_2$ are free parameters. In order to extend the formulation outside the shallow water domain, these authors replaced \citeauthor{Thornton1983}'s \citeyear{Thornton1983} breaking criterion based on the wave height ($H$) to water depth ($h$) ratio ($\gamma=H/h=0.42$) with an adaptation of \citeauthor{Miche1944b}'s \citeyear{Miche1944b} wave breaking parameter:

\begin{equation}
	\beta_r = \frac{\overline{k_r}(f_c)H_r(f_c)}{\tanh{(\overline{k_r}(f_c)h)}} \label{eq:02:02:05}
\end{equation}

\noindent in which

\begin{equation}
	\overline{k_r}(f_c) = \frac{\int_{0}^{\infty} U_{f_c}(f) k (f) E(f)df }{\int_{0}^{\infty} U_{f_c}(f)E(f)df} \label{eq:02:02:06}
\end{equation}

\noindent and

\begin{equation}
	\tilde{\beta} = b  (b_3\tanh(kh)^3 -b_2\tanh(kh)^2 + b_1\tanh(kh) - b_0) \label{eq:nested:model}
\end{equation}

\noindent in which $b=0.48$, $b_3=1.0314$, $b_2= 1.9958$, $b_1=1.5522$, and $b_0= 0.1885$. In their model, the variable $\tilde{\beta}$ was obtained via numerical calculations of regular nearly breaking waves using the stream wave theory of \citeA{Dean1965}. Finally, the wave breaking probability is obtained as:

\begin{equation}
	P_b(f_c) = \int_{0}^{\infty}P(H, f_c)W(H, f_c)dH \le 1. \label{eq:02:02:08}
\end{equation}

\noindent To keep consistency with Section \ref{sec:parametric:models:banner:2002}, $P_b$ will be only considered at the spectral peak; other definitions are, however, also possible.

\subsection{Models based on  Phillips’ (1985) $\Lambda(c)$}\label{sec:parametric:models:philips:1985}

The major issue with the previous models is the difficulty to obtain reliable observations of the wave breaking probabilities as a spectral distribution solely from point measurements. Due to the presence of different wave scales at the time and location, it is indeed difficult to assign the breaking occurrence to a given wave frequency of wave number. To avoid this problem, \citeA{Phillips1985} proposed to use the speed of the breaking front as a proxy for the phase speed of the carrying wave. 
 \citeA{Phillips1985} defined the parameter $\Lambda(c)dc$ as the \textquote{average total length per unit surface area of breaking fronts that have velocities in the range $c$ to $c+dc$} and then defined the following quantities:

\begin{equation}
	L = \int \Lambda(c) dc \label{eq:02:03:01}
\end{equation}

\noindent and

\begin{equation}
	R = \int c \Lambda(c) dc \label{eq:02:03:02}
\end{equation}

\noindent which represent the \textquote{total length of breaking fronts per unit area} (Equation \ref{eq:02:03:01}) and \textquote{the total number of breaking waves of all scales passing a given point per unit time} (Equation \ref{eq:02:03:02}). Assuming that \citeA{Phillips1985} assumptions hold, it is possible to obtain parametric models for $\Lambda$ from known variables (e.g., wind speed) and, consequently, for $P_b$ (see Equation \ref{eq:pb:from:lambda}). 

\subsubsection{Melville and Matusov (2002)}\label{sec:parametric:models:melville:2002}

\citeauthor{Melville2002}'s \citeyear{Melville2002} model for $\Lambda(c)$ relies only on the wind speed measured at 10m ($U_{10}$) to obtain $\Lambda(c)$. Following \citeA{Melville2002} and using the explicit formula given by \citeA{Reul2003}, this parameterization is written as:

\begin{equation}
	\Lambda(c) = p_1 \left[\frac{U_{10}}{10} \right]^3 10^{-4} \exp{[-(p_2c)]} \label{eq:02:03:01:01}
\end{equation}

\noindent in which $p_1$ and $p_2$ are constants. For their data, \citeA{Melville2002} found $p_1=3.3$ and $p_2=0.64$. As discussed by \citeA{Reul2003}, this formulation approaches \citeauthor{Phillips1985}'s \citeyear{Phillips1985} theoretical $c^{-6}$ but may overly estimates the amount of small breakers.
	
\subsubsection{Sutherland and Melville (2013)}\label{sec:parametric:models:sutherland:2002}

\citeA{Sutherland2013} used dimensional analysis to scale $\Lambda(c$) and obtain a parameterization that is a function of the wind drag ($u_*$), peak wave phase speed ($c_p$), significant wave height ($H_s$) and three constants. From \citeauthor{Sutherland2013}'s \citeyear{Sutherland2013} Equation 9 and their Figure 4, $\Lambda(c$) is calculated as:
	
\begin{equation}
	\Lambda(c) = p_1 \frac{g}{c_{p}^3}\left(\frac{u_*}{c_{p}} \right)^{p_2} \left(\frac{c}{\sqrt{gH_s}}\left(\frac{gH_s}{c_{p}^2}\right)^{p_3}\right)^{-6} \label{eq:02:03:02:01}
\end{equation}

\noindent where $p_1=0.05$, $p_2=0.5$, and $p_3=0.1$ are constants obtained from the available data. Their formulation reproduces \citeauthor{Phillips1985}'s \citeyear{Phillips1985} $c^{-6}$ frequency dependency but does not have the typical roll-off at low $c$ as these authors chose to use infrared (other than visible) imagery to obtain and model their $\Lambda(c)$. This choice included the contribution of micro-scale breakers that do generate visible bubbles in their model, hence the difference.
	
\subsubsection{Romero (2019)}\label{sec:parametric:models:romero:2019}

Recently, \citeA{Romero2019} developed and implemented a new wave breaking parameterization in WaveWatchIII which relies exclusively on $\Lambda(c)$. Differently from previous parameterizations,  \citeauthor{Romero2019}'s \citeyear{Romero2019} takes into account both the modulations due to winds and long waves on $\Lambda(c)$. His model is fairly general but depends on six free parameters that needed to be laboriously obtained by comparing WaveWatchIII's significant wave height outputs with available measured significant wave heights from buoy data. In \citeauthor{Romero2019}'s \citeyear{Romero2019} model, $\Lambda$ was modelled assuming that it is proportional to the crest lengths exceeding a slope threshold:

\begin{equation}
	\Lambda(f, \theta) = \left( \frac{2(2\pi)^2p_1}{g} \right)f \exp{\left[-\left( \frac{p_2}{B(f, \theta)}\right) \right] M_{LW} M_{W}} \label{eq:02:03:03:01}
\end{equation}
	
\noindent where $p_1=3.5 \mathsf{x} 10^{-5}$ and $p_2=5 \mathsf{x}  10^{-3}$ are constants to be obtained from the data, $M_{LW}$ is the modulation due to long waves, $M_W$ is the modulation due to winds and $B(f, \theta)$  is the directional wave breaking saturation spectra:
	
\begin{equation}
	B(f) = \int_{0}^{2\pi}B(f, \theta) d\theta = E(f)\left(\frac{2\pi f^5}{2g} \right). \label{eq:02:03:03:02}
\end{equation}

\noindent The modulation due to long waves is calculated according to \citeA{Guimaraes2018t}:
	
\begin{equation}
	M_{LW} = \left[ 1 + p_3\sqrt{\cmss{(E(f))}} \cos^2(\theta-\hat{\theta}) \right]^{p_4}  \label{eq:02:03:03:03}
\end{equation}
	
\noindent where $p_3=400$ and $p_4=3/2$ are also best-fit constants found by \citeA{Romero2019}. The cumulative mean square slope ($\cmss$) is defined as:
	
\begin{equation}
	\cmss = \int_{0}^{\infty} E(f) \left( \frac{(2\pi)^4f^4}{g^2} \right)df. \label{eq:02:03:03:05}
\end{equation}

\noindent and

\begin{equation}
    \hat{\theta} = \tan \left( \frac{\int E(f, \theta) \sin{(\theta)} dfd\theta} {\int E(f, \theta) \cos{(\theta)} dfd\theta} \right)
\end{equation}

\noindent The modulation due to the wind is computed as:
	
\begin{equation}
	M_W = \frac{\left( 1 + p_5 \max{\left( 1, \frac{f}{f_0} \right)} \right)}{\left(1 + p_5 \right)} \label{eq:02:03:03:06}
\end{equation}

\noindent with

\begin{equation}
    f_0={p_6 \frac{1}{u_{*}} \frac{g}{2\pi}}    
\end{equation}

\noindent where $p_5=0.9$ is a constant related to the DIA algorithm and  $p_6 = 3/28$ is yet another constant. Finally, the conversion from $\Lambda(f)$ to $\Lambda(c)$ is done using the relation $\Lambda(c)dc = \Lambda(f)df$ and the linear dispersion relation (see \citeauthor{Romero2019}'s \citeyear{Romero2019} Eqs. 17-23 for details).

\acknowledgments

This work benefited from France Energies Marines and State financing managed by
the National Research Agency under the Investments for the Future program bearing the reference numbers ANR-10-IED-0006-14 and ANR-10-IEED-0006-26 for the projects DiME and CARAVELE. The authors' thank Peter Sutherland and Jim Thompson for kindly sharing their data.

\section*{Data Availability}

All data used in this publication has been previously published by \citeA{Banner2000}, \citeA{Sutherland2013}, \citeA{Schwendeman2014}.

\bibliography{library.bib}

\end{document}